\documentclass[a4paper,fleqn,usenatbib]{mnras}

\usepackage[T1]{fontenc}
\usepackage{ae,aecompl}

\usepackage{graphicx}	% Including figure files
\usepackage{amsmath}	% Advanced maths commands
\usepackage{amssymb}	% Extra maths symbols
\usepackage{multirow}
\usepackage{color}
\usepackage{ulem}
\title[Observation of polarised hard X-ray emission from the Crab]{Observation of polarised hard X-ray emission from the Crab by the {\it PoGOLite Pathfinder}}

\author[M. Chauvin et al.]{
M. Chauvin,$^{1,2}$
H.-G. Flor\'{e}n$^{3}$, 
M. Jackson$^{1}$, 
T. Kamae$^{4}$, 
T. Kawano$^{5}$, 
\newauthor
M. Kiss$^{1,2}$, 
M. Kole$^{1,2}$, 
V. Mikhalev$^{1,2}$, 
E. Moretti$^{1,2,6}$, 
G. Olofsson$^{3}$, 
\newauthor
S. Rydstr\"{o}m$^{1,2}$, 
H. Takahashi$^{5}$, 
A. Iyudin$^{7}$,
M. Arimoto$^{8}$,  
Y. Fukazawa$^{5}$, 
J. Kataoka$^{9}$, 
\newauthor
N. Kawai$^{8}$, 
T. Mizuno$^{5}$, 
F. Ryde$^{1,2}$, 
H. Tajima$^{10}$, 
T. Takahashi$^{11}$, 
\newauthor
M. Pearce$^{1,2}$\thanks{E-mail: pearce@kth.se}
\\
$^{1}$KTH Royal Institute of Technology, Department of Physics, 106 91 Stockholm, Sweden\\
$^{2}$The Oskar Klein Centre for Cosmoparticle Physics, AlbaNova University Centre, 106 91 Stockholm, Sweden\\
$^{3}$Stockholm University, Department of Astronomy, 106 91 Stockholm, Sweden\\
$^{4}$University of Tokyo, Department of Physics, 113-0033 Tokyo, Japan\\
$^{5}$Hiroshima University, Department of Physical Science, Hiroshima 739-8526, Japan\\
$^{6}$Max Planck Institute for Physics, 80805 Munich, Germany\\
$^{7}$Skobeltsyn Institute of Nuclear Physics, Moscow State University by M.V. Lomonosov, 119991 Moscow, Russia\\
$^{8}$Department of Physics, Tokyo Institute of Technology, Tokyo 152-8551, Japan\\
$^{9}$Research Institute for Science and Engineering, Waseda University, Tokyo 169-8555, Japan\\
$^{10}$Institute for Space-Earth Environment Research, Nagoya University, Aichi 464-8601, Japan\\
$^{11}$Institute of Space and Astronautical Science, Japan Aerospace Exploration Agency, Kanagawa 252-5210, Japan
}

\date{Accepted 9 November 2015. Received 6 November 2015; in original form 2 September 2015}

\pubyear{2015}

\begin{document}
\label{firstpage}
\pagerange{\pageref{firstpage}--\pageref{lastpage}}
\maketitle

\begin{abstract}
We have measured the linear polarisation of hard X-ray emission from the Crab in a previously unexplored energy interval, 20--120~keV. 
{\color{black} The introduction of two new observational parameters, the polarisation fraction and angle stands to disentangle geometrical and physical effects, 
thereby providing information on the pulsar wind geometry and magnetic field environment.} 
Measurements are conducted using the {\it PoGOLite Pathfinder} - a balloon-borne polarimeter. 
{\color{black} Polarisation is determined by measuring the azimuthal Compton scattering angle of incident X-rays in an array of plastic scintillators housed in an anticoincidence well.}
The polarimetric response has been characterised prior to flight using both polarised and unpolarised calibration sources. 
We address possible systematic effects through observations of a background field. 
The measured polarisation fraction for the integrated Crab light-curve is ($18.4^{+9.8}_{-10.6}$)\%, corresponding to an upper limit (99\% credibility) of 42.4\%, for a polarisation angle of ($149.2\pm16.0)^\circ$. 
\end{abstract}


%%%%%%%%%%%%%%%%%%%%%%%%%%%%%%%%%%%%%%%%%%%%%%%%%%

%%%%%%%%%%%%%%%%% BODY OF PAPER %%%%%%%%%%%%%%%%%%
\section{Introduction}

At radio, optical and infrared wavelengths, polarimetry has become an established probe of source radiation mechanisms and geometry. 
The application to X-ray sources has not evolved as rapidly since relatively long exposures of sensitive, purpose-built instruments are needed. The positive definite nature of measurements presents a significant challenge for instrument design and characterisation. 
Measurements of linear X-ray polarisation provide two additional observation parameters, the polarisation fraction (\%) and polarisation angle (photon electric vector orientation). The addition of these parameters stands to elucidate emission processes and geometry in sources containing ordered magnetic fields, e.g. pulsars, or high degrees of asymmetry, e.g. accretion disks and columns~\citep[e.g.][]{Lei1997, Krawczynski2011}. 

The scientific potential of X-ray polarisation studies is recognised by the plethora of proposed missions\footnote{For a recent survey, see http://ttt.astro.su.se/groups/head/cost14/}. Eventually, satellite-based polarimeters are expected to provide defining studies of a large number of sources, e.g. {\it XIPE}\mbox{~\citep[][]{Soffitta2013}} (ESA Medium Class mission), {\it IXPE}~\citep[][]{Weisskopf2008} and {\it PRAXyS} (NASA Small Explorer missions) were recently selected for further study.
Despite comparatively short observation times and the effect of residual atmosphere, polarisation measurements using purpose-built instruments operating from stratospheric  ballooning platforms 
(altitude $\sim$40~km) can provide initial data-sets and opportunities for methodology development. 
In this paper, we report polarisation results from such a balloon mission - the {\it PoGOLite Pathfinder}~\citep[][]{Chauvin2015a}, which is a 
purpose-designed Compton polarimeter with sensitivity in the energy range 20--240~keV. The polarimetric response has been characterised prior to flight with both polarised and unpolarised beams~\citep[][]{Chauvin2016}.

As one of the brightest sources on the sky, the Crab pulsar and wind nebula system is a natural first light candidate for {\it PoGOLite}.
Although perhaps the most studied extrasolar celestial object~\citep[][]{Hester2008}, details of the high energy emission, including fundamental questions such as location, remain unresolved. 
The emission site may 
(a) take place in open field line regions close to the neutron star surface above the polar cap, as seen for radio pulses~\citep[][]{Harding1998, Daugherty1996};
(b) occur in the outer magnetosphere along the last open field line between the null-charge surface and the light cylinder~\citep[][]{Romani1996, Cheng2000}; 
(c) or concern an intermediate region~\citep[][]{Dyks2004}, along the edge of the open field region extending from the surface of the neutron star out to the light cylinder.
The models predict high degrees of linear polarisation with significant variations across the pulsar phase. This mirrors the behaviour seen in optical measurements~\citep[][]{Slowikowska2009} due to a common dependence on field line geometry.
Precision spectral studies of the Crab by {\it Fermi}~\citep[][]{Abdo2010} favour high energy emission in the outer magnetosphere.
The emission is almost certainly synchrotron in nature, but curvature radiation has also been proposed 
by~\citet{Sturrock1975}.
An observation of the spectral properties of the emission is not sufficient to distinguish between these two processes. Polarisation measurements provide discrimination since the polarisation vector in
synchrotron emission is perpendicular to the magnetic field line direction while there is a parallel arrangement for curvature radiation.  
{\color{black} Furthermore, \citet{Satori1967} have proposed a two temperature plasma model describing Crab emissions through bremsstrahlung processes. In this case, the emission is predicted to be unpolarised.}

\citet{Novick1972} reported the first detection of X-ray polarisation (5-20~keV) for the Crab nebula with a sounding rocket payload.
A highly significant measurement was subsequently provided 
by an instrument on-board the {\it OSO-8} satellite mission~\citep[][]{Weisskopf1976} during a 256 ks observation.
{\color{black} A} high polarisation fraction {\color{black} was} reported, {\color{black} e.g.} (19.2$\pm$1.0)\% at 2.6~keV, {\color{black} compatible with expectations for} 
{\color{black}} synchrotron processes {\color{black}}, as expected from earlier spectral studies at radio and optical wavelengths.   
The measured polarisation angle of (156.4$\pm$1.4)$^\circ$ is displaced from the direction of the pulsar spin axis projected onto the sky, (124{\color{black}.0}$\pm$0.1)$^\circ$,
{\color{black} inferred from the torus inclination}~\citep[][]{Ng2004}.  
More recently, the IBIS~\citep[][]{Forot2008} and SPI~\citep[][]{Dean2008} instruments on-board the {\it INTEGRAL} satellite have measured the polarisation of Crab emissions for selected
pulsar phase windows. 
Unlike {\it OSO-8}, these instruments were not designed for polarimetry nor calibrated as such prior to launch. 

%%%%%%%%%%%%%%%%%%%%%%%%%%%%%%%%%%%%%%%%%%%%%%%%%%
\section{Method}
We perform observations using {\it PoGOLite}, a Compton polarimeter.
The Klein Nishina differential Compton scattering cross-section dictates that X-rays are more likely to scatter in the direction perpendicular to the polarisation vector. The resulting sinusoidal distribution, $f(\phi)$, of azimuthal scattering angles, $\phi$, is referred to as a modulation curve and carries information about the polarisation fraction and angle. 
The modulation curve has the form 
\begin{equation}
f(\phi)=N(1 + A \cos(2(\phi-\phi_0))),
\label{eq:mod_curve}
\end{equation}
where $A$ is the peak-to-peak amplitude of the modulation, $N$ is the number of events and 
$\phi_0$ is the phase of the modulation, related to the polarisation angle, $\psi = \phi_0-90^\circ$. 

Azimuthal scattering angles are measured in a close-packed array of 20~cm long and 3~cm wide (center-to-center) hexagonal plastic scintillator rods. Two-hit polarisation events are characterised by a Compton scatter in one rod followed by a second Compton scatter or a photoelectric absorption in another rod. The entire detector volume is rotated around the viewing axis once every 300~s in order to mitigate instrument systematics such as variations in efficiency between detector elements. 
Aperture background is reduced with an active collimator system which defines the field-of-view of the polarimeter to $2.4^\circ \times 2.6^\circ$.
The plastic scintillator target is housed within a segmented active BGO ({Bi$_{4}$Ge$_{3}$O$_{12}$}) anticoincidence well, surrounded by a thick polyethylene neutron shield.

Assuming an unpolarised background during observations, the modulation factor is defined as 
\begin{equation}
M=A\times(1+1/{{\cal{R}}}), 
\end{equation}
where ${\cal{R}}$ is the signal-to-background ratio. The reconstructed polarisation fraction is defined as $p=M/M_{100}$, where $M_{100}$ is the modulation factor for a 100\% polarised incident beam with defined spectral properties. When performing a $\chi^2$ fit to a histogram of scattering angles the statistical uncertainty on $p$ is derived by standard error propagation as 
\begin{equation}
\begin{split}
\sigma_{p}^2=\frac{1}{M_{100}^2}\Bigg[\bigg(1+\frac{1}{{\cal{R}}}\bigg)^2\times\sigma^2_{A} +\frac{A^2}{{\cal{R}}^4}\times\sigma_{{\cal{R}}}^2+\\
\frac{A^2\times\bigg(1+\frac{1}{{\cal{R}}}\bigg)^2}{M_{100}^2}\times\sigma_{M_{100}}^2\Bigg]
\end{split}
\end{equation}
where $\sigma_{A}$, $\sigma_{{\cal{R}}}$ and $\sigma_{M_{100}}$ are the statistical uncertainties on $A$, ${\cal{R}}$ and $M_{100}$, respectively. The statistical uncertainty on the polarisation angle is simply $\sigma_{\psi}=\sigma_{\phi_0}$, as derived from the $\chi^2$ fit.

An alternative method for computing $p$ uses the Stokes parameters. Following the formalism in~\citet{Kislat2015},
$Q$ and $U$ are defined as
\begin{equation}
%\begin{split}
Q=\frac{1}{N_S}\sum\limits_{i=1}^N\cos2\psi_i\,,
U=\frac{1}{N_S}\sum\limits_{i=1}^N\sin2\psi_i, 
%\end{split}
\end{equation}
where $N_S$ is the total number of signal events. This method has the advantage that it is 
not affected by subjectivity when choosing the binning of the modulation curve. However, the absence of goodness-of-fit information, which can reveal distortions of the modulation curve, may yield spurious polarisation results. The polarisation fraction and angle are given by
\begin{equation}
\begin{split}
p=\frac{2}{M_{100}}\sqrt{Q^2+U^2}\,, 
\psi=\frac{1}{2}\arctan{\frac{U}{Q}}.
\end{split}
\label{eq:Stokes_param}
\end{equation}
The statistical uncertainties on these quantities are
\begin{equation}
\begin{split}
\sigma_{p}\approx\sqrt{\frac{2-p^2M_{100}^2}{(N-1)M_{100}^2}}\,,
\sigma_{\psi}\approx\frac{1}{pM_{100}\sqrt{2(N-1)}},
\end{split}
\label{eq:Stokes_uncertainties}
\end{equation}
not taking into account the statistical uncertainty on ${\cal{R}}$ and $M_{100}$.

A figure-of-merit often used for describing the sensitivity of a polarimeter is the Minimum Detectable Polarisation, MDP ~\citep[][]{Weisskopf2010}, where 
\begin{equation}
\textrm{MDP}=\frac{4.29\sqrt{N}}{M_{100}N_{S}}
\end{equation}
at 99\% confidence level.
If a measurement yields a polarisation fraction equal to the MDP, there is a 1\% probability that the measured value arises from a statistical fluctuation of an unpolarised flux.
A measurement of polarisation is possible when the reconstructed polarisation fraction lies below the MDP but the uncertainties on such a measurement are non-Gaussian leading to a biased polarisation fraction, i.e. $\Pi\neq p$, where $\Pi$ is the true polarisation fraction. Although $\Psi=\psi$, where $\Psi$ is the true polarisation angle, $\sigma_\Psi\neq\sigma_\psi$ 
($\sigma_\Psi$ is the uncertainty on the true polarisation angle)
because the uncertainty given by equation~\ref{eq:Stokes_uncertainties} or by a $\chi^2$ fit to the modulation curve is assumed to be Gaussian but $\Psi$ is not Gaussian distributed. 
We have therefore developed a Bayesian method~\citep[][]{Mikhalev2015} based on~\citet{Maier2014} for converting $p$ to $\Pi$ and computing the two dimensional credibility regions for $\Pi$ and $\Psi$. 
{\color{black}
The likelihood is sampled using a Monte Carlo approach and inverted using the integral form of Bayes' theorem. This allows the posterior distribution of ($\Pi$,$\Psi$) to be determined for a measured polarisation $p$. This Bayesian approach requires a prior, which, considering the discussion in~\cite{Quinn2012}, is chosen to be uniform in polar coordinates ($\Pi$,$\Psi$) and limited to $0<\Pi<1$.}

%%%%%%%%%%%%%%%%%%%%%%%%%%%%%%%%%%%%%%%%%%%%%%%%%%%%%%%%%%%%%%%%%%%%%%
\section{Observations and data reduction}
\label{sec:observations}
The {\it PoGOLite Pathfinder} was launched from the Esrange Space Centre in Northern Sweden on July 12th 2013 on a near circumpolar flight~\citep[][]{Chauvin2015a}. The polarimeter is suspended $\sim$100~m under a helium filled 10$^6$~m$^3$ balloon. An attitude control system aligns the polarimeter optical axis to sources with a precision exceeding $0.01^\circ$ in RA and Dec. We observed the Crab during each of the first three days of the flight for a total of 32.9~ks. 
Observing conditions are summarised in Table~\ref{table:observations}. 
We only consider observations with column density $<$6.8~g cm$^{-2}$ to avoid degrading the MDP. A total of 122\,993 two-hit polarisation events are considered for analysis for the energy range 20--120~keV, which is dictated by the polarimeter effective area for Crab observations and atmospheric attenuation. The average energy of measured photons is 46~keV.
The procedure described in~\citep[][]{Chauvin2016} was followed to compute the Crab $M_{100}$ for the flight event selection criteria described in~\citep[][]{Chauvin2015a}, 
yielding $(21.4\pm1.5)\%$.
We used data\footnote{Public data-set 104001070.} from the HXD instrument~\citep[][]{Takahashi2007} on the {\it Suzaku} satellite
(10-70~keV energy range) to produce a reference light-curve. 
The difference in energy range between {\it PoGOLite} and HXD has negligible impact on the reference light-curve~\citep[][]{Kokubun2007}. 
We compared the measured and reference light-curves to determine ${\cal{R}}=0.252\pm0.026$ which results in $\textrm{MDP}=(28.4\pm2.2)\%$. 

\begin{table}
\centering
\caption{\label{observations}Summary of Crab observations.}
\label{table:observations}
\begin{tabular}{|l|c|c|c|r|}
\hline
{Date} & {Observation} & {Altitude} & {Column} &  {Two-hit} \\
{in 2013} & {time} & {   } & {density} &  {events} \\
{ } & {(ks)} & {(km)} & {(g cm$^{-2}$)} &  {} \\
\hline
July 12th & 2.6 & 38.6-39.9 & 5.9-6.8 & 11\,154 \\
\hline
July 13th & 19.1 & 38.3-39.9 & 4.6-6.8 & 70\,700 \\
\hline
July 14th & 11.2 & 38.2-39.0 &  5.7-6.8 & 41\,139 \\
%\hline
%Total & 549 & 38.2-39.9 & 4.6-6.8 & 122993 \\
\hline
\end{tabular}
\end{table}

Polarisation fraction is a positive definite quantity. 
It is therefore important to show that observations of background fields yield 
a modulation curve which is statistically consistent with an unpolarised signal. 
Prior to flight, we planned observations of background fields displaced 5$^\circ$ in RA and Dec from the Crab. Problems with the thermal control of the polarimeter during observations 
prevented this approach. 
Instead, we conducted background studies in the anti-Sun direction with the polarimeter directed at Cygnus~X-1,
which was originally foreseen as a secondary observation target. 
However, the prevailing soft spectral state and average column density during background observations of 6.5 g cm$^{-2}$ (5.7 g cm$^{-2}$ for Crab observations) resulted in an estimated two-hit ${\cal{R}}$ of 1:27 for the {\it PoGOLite} energy range, based on fluxes measured by {\it Swift} BAT~\citep[][]{Krimm2013}. This results in a negligible signal count rate.

%
%%%%%%%%%%%%%%%%%%%%%%%%%%%%%%%%%%%%%%%%%%%%%%%%%%%%%%%%%%%%%%%%%%%%%
\section{Analysis and results}
\label{sec:results}
%
%%%%%%%%%%%%%%%%%%%%%%%%%%%%
\subsection{Background observations}
\begin{figure}
 \centering
    \includegraphics[width=\columnwidth]{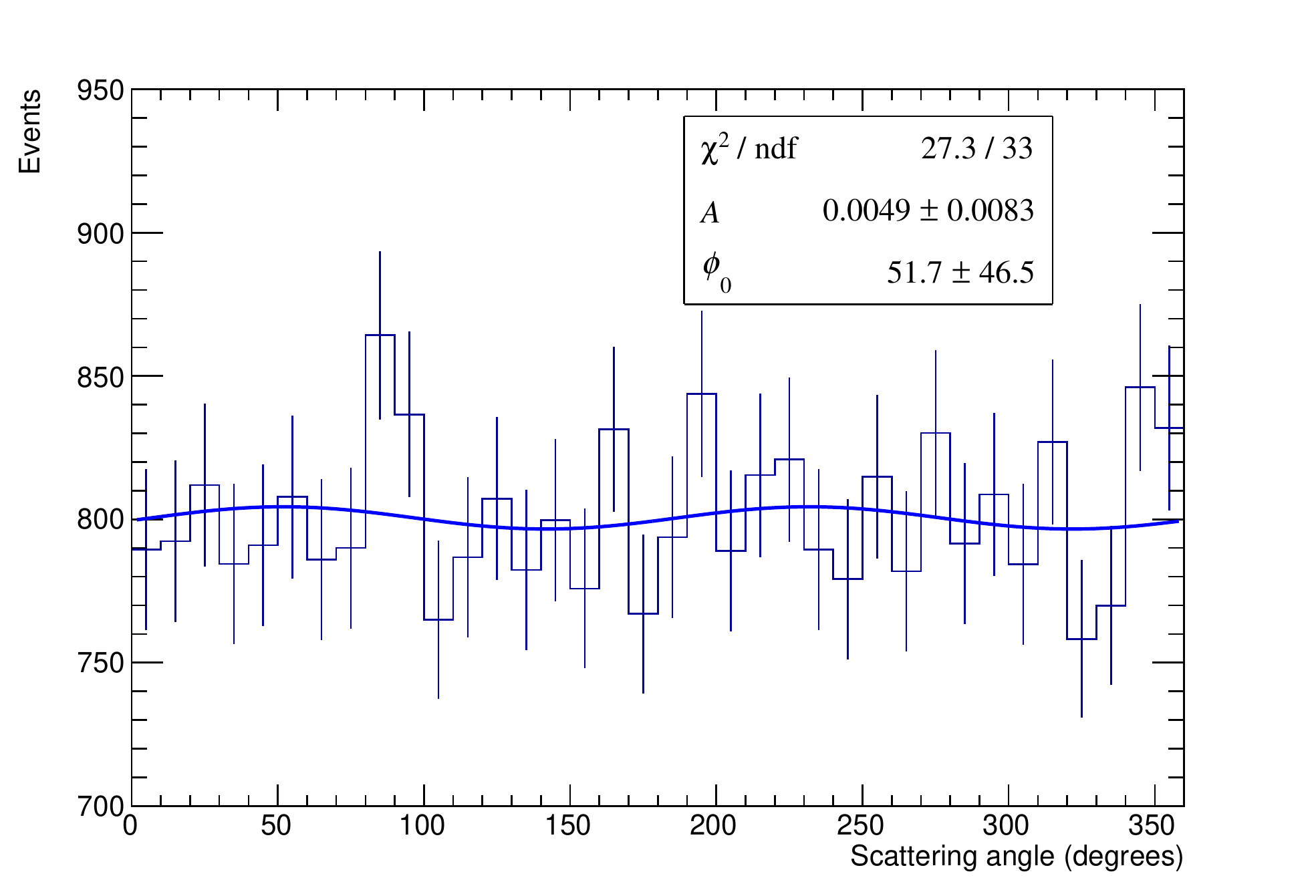}
\caption{The modulation curve for background observations.}
  \label{fig:bmodcurve}
\end{figure}

As described in~\citet{Chauvin2016}, anisotropic background may result in a modulation of the scattering angle, which cannot be removed by the rotation of the polarimeter around its optical axis. 
To show that atmospheric background does not induce a false modulation signature, we use background observations (9~ks) with a similar elevation range as Crab observations. Each scattering angle is transformed to account for the instantaneous rotation of the polarimeter and inclination of the gondola. The resulting modulation curve is shown in Figure~\ref{fig:bmodcurve} and the fitted modulation parameters are summarised in 
Table~\ref{tab:pol_param}. There is no statistical evidence for background-induced modulation. 

\begin{table}
\centering
\caption{Polarisation parameters as derived from $\chi^2$ fitting and Stokes method. Here $A$, $p$ and $\psi$ are the modulation amplitude, polarisation fraction and polarisation angle, respectively.}
\label{tab:pol_param}
\begin{tabular}{cc|c|c|c|c|l}
\cline{3-5}
& & {$A$} & {$p$} & {$\psi$} \\ \cline{1-5}
\multicolumn{1}{ |c  }{\multirow{2}{*}{\bf Bkgnd} } &
\multicolumn{1}{ |c| }{$\chi^2$} & $(0.49\pm0.83)\%$ & - & -     \\ \cline{2-5}
\multicolumn{1}{ |c  }{}                        &
\multicolumn{1}{ |c| }{Stokes} & $(0.49\pm0.83)\%$ & - & -    \\ \cline{1-5}
\multicolumn{1}{ |c  }{\multirow{2}{*}{\bf Crab} } &
\multicolumn{1}{ |c| }{$\chi^2$} & $(0.91\pm0.40)\%$ & $(21.0\pm9.6)\%$ & $(150\pm13)^\circ$ \\ \cline{2-5}
\multicolumn{1}{ |c  }{}                        &
\multicolumn{1}{ |c| }{Stokes} & $(0.90\pm0.40)\%$ & $(20.8\pm9.4)\%$ & $(149\pm13)^\circ$ \\ \cline{1-5}
\end{tabular}
\end{table}
%
%%%%%%%%%%%%%%%%%%%%%%%%%%%%
\subsection{Crab observations}
\begin{figure}
 \centering
    \includegraphics[width=\columnwidth]{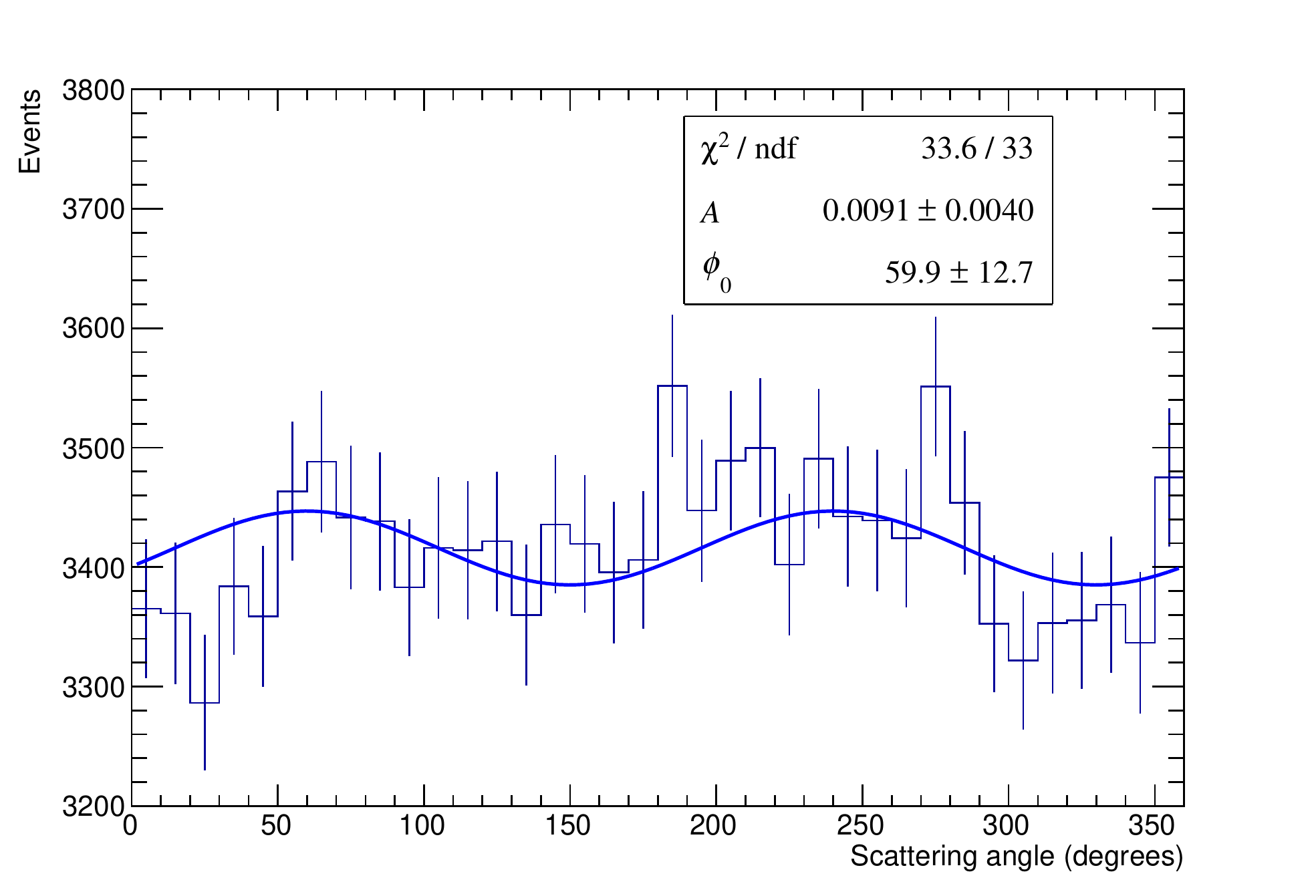}
\caption{The modulation curve for Crab observations.}
  \label{fig:crabmodcurve}
\end{figure}

Scattering angles were transformed to the Crab frame of reference by applying a field rotation correction in the range $-20^\circ$~to~$+24^\circ$. Without this correction the modulation amplitude was $A=(0.59\pm0.40)\%$.
With the correction included the amplitude is more significant, $A=(0.91\pm0.40)\%$, as shown in Figure~\ref{fig:crabmodcurve}. The polarisation parameters are shown in Table~\ref{tab:pol_param}. 
Considering data from 13th and 14th July separately yields statistically consistent results.
Figure~\ref{fig:crabmodcurve} is primarily shown to demonstrate goodness-of-fit. The Stokes method avoids binning subjectivity and yields $p=(20.8\pm9.4)\%$ and $\psi=(149\pm13)^\circ${\color{black}, where the modest significance reflects that $p < \mathrm{MDP}$.}  

%
%%%%%%%%%%%%%%%%%%%%%%%%%%%%
\subsection{Crab polarisation fraction and angle}
\begin{figure}
 \centering
    \includegraphics[width=\columnwidth]{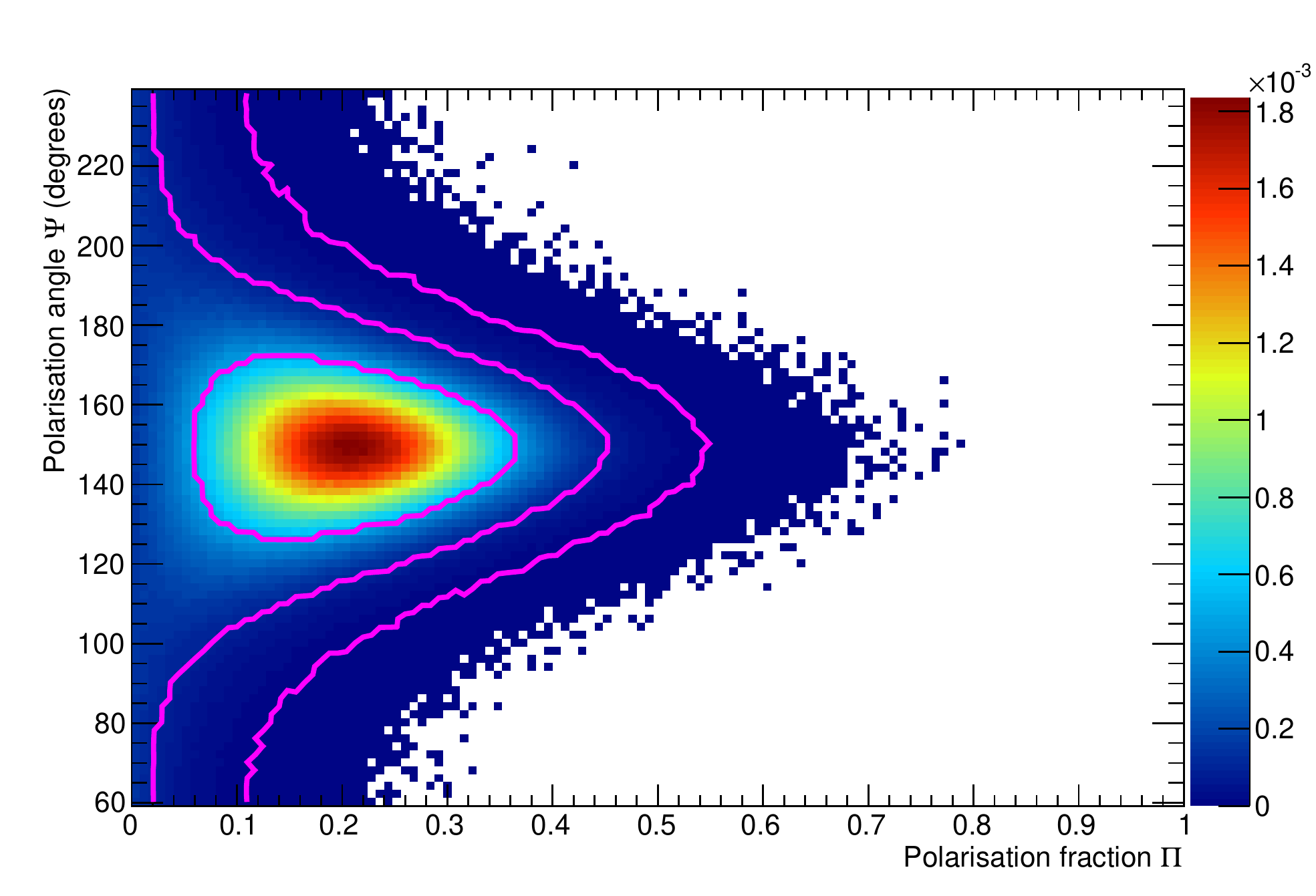}
\caption{The posterior distribution of the polarisation fraction and angle from {\it PoGOLite} Crab observations. The magenta lines are contours corresponding to 1, 2 and 3 standard deviation Gaussian probability content. The color bar is the probability per bin.}
  \label{fig:contour}
\end{figure}

{\color{black} Since $p < \mathrm{MDP}$, a Bayesian approach is required to properly account for statistical uncertainties and to determine credibility regions when obtaining the true polarisation fraction, $\Pi$, and angle, $\Psi$, from the reconstructed polarisation fraction, $p$, and angle, $\psi$, in the Stokes analysis. }
As described in~\citet{Mikhalev2015}, this approach also accounts for the uncertainty on the measurement of ${\cal{R}}$ due to the fitting of the reference light-curve and the uncertainty on $M_{100}$ due to limited calibration statistics. 
The resulting probability distribution contour plot is shown in Figure~\ref{fig:contour}.

\begin{figure}
 \centering
 \includegraphics[width=\columnwidth]{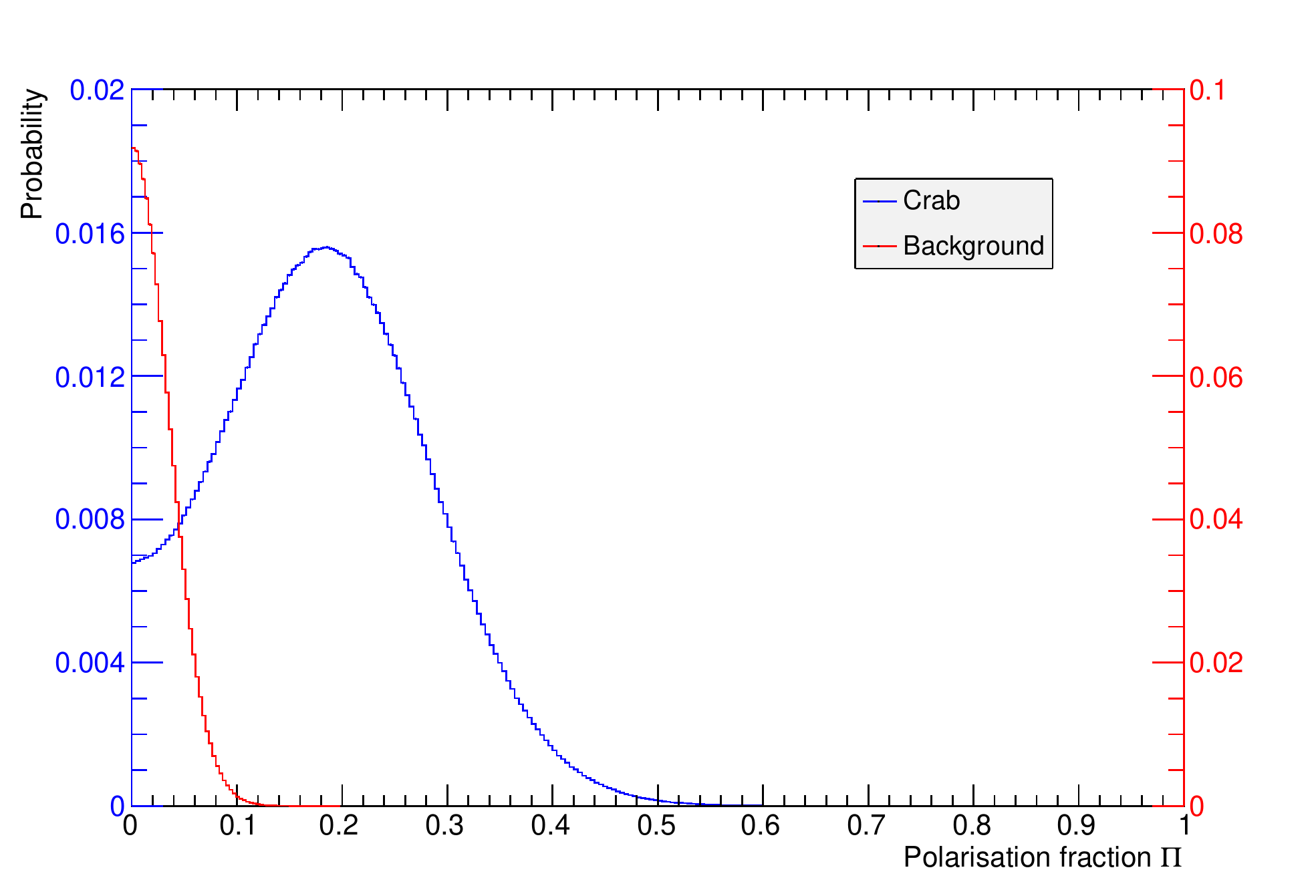}
%\caption{The polarisation fraction marginalised over the polarisation angle for Crab observations.}
\caption{{\color{black} The polarisation fraction marginalised over the polarisation angle for Crab observations (blue, left vertical scale) and the background contribution derived from background observations (red, right vertical scale).}}
  \label{fig:margpol}
\end{figure}

\begin{figure}
 \centering
    \includegraphics[width=\columnwidth]{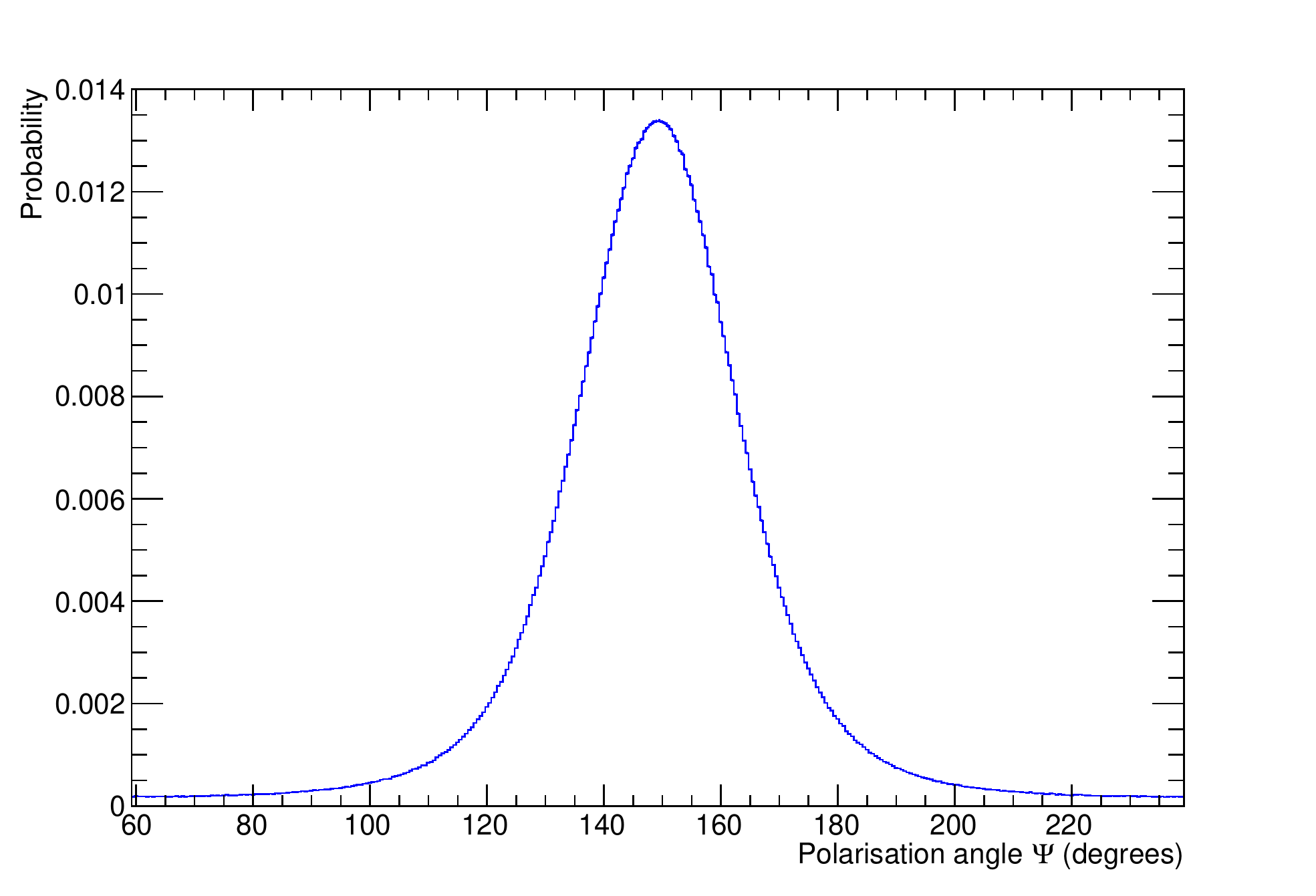}
\caption{The polarisation angle marginalised over the polarisation fraction for Crab observations.}
  \label{fig:margangle}
\end{figure}

Figure~\ref{fig:margpol} is obtained by marginalising the contour plot over the polarisation angle (i.e. projecting onto the polarisation fraction axis). The peak corresponds to the most likely (highest posterior density) true polarisation $\Pi=(18.4^{+9.8}_{-10.6})\%$ where the asymmetric uncertainty represents the tightest credibility region (not the statistical uncertainty on the peak) enclosing $68.3\%$ (one Gaussian standard deviation) probability content. The upper limit, at 99\% credibility, is $42.4\%$. Marginalising over the polarisation fraction gives Figure~\ref{fig:margangle}. The polarisation angle is $\Psi=(149.2\pm16.0)^\circ$, measured relative to North in an anticlockwise direction (i.e. to the East). 
{\color{black} 
The posterior distribution for the background contribution derived from background observations is also shown in Figure~\ref{fig:margpol}. The corresponding point-estimate for the polarisation fraction is 0\%. The 99\% upper limit on the background contribution is 11.2\%.
}

%%%%%%%%%%%%%%%%%%%%%%%%%%%%%%%%%%%%%%%%%%%%%%%%%%
\section{Discussion}

In this paper, we present the first measurement of the linear polarisation of Crab emissions in the 20-120~keV energy interval. 
We have used a purpose-built polarimeter which has been calibrated with both 
polarised and unpolarised beams prior to launch, as required due to the positive definite nature of the measurements. 
Through the use of background field observations, we have addressed the response of the polarimeter to potentially anisotropic background, and show that this does not induce a statistically significant polarisation signal.
We present results integrated over the entire Crab light-curve owing to the relatively short 
observation time (33~ks) due to a technical failure during the third day of the circumpolar stratospheric balloon flight.
Due to the relatively large field-of-view of the polarimeter, contributions from the nebula and pulsar can only be obtained by making phase selections on the light-curve.
For the most favourable case of the nebula, the required phase window of 0.4 pulsation periods rejects $\sim 65\%$ of the two-hit events and raises the MDP to $51\%$. 
The approach is therefore not feasible given the size of current data-sets.  

{\color{black} Our results are compatible with those reported at lower energies by~\citet{Weisskopf1976} for the Crab nebula. This indicates that the phase-integrated contribution from the pulsar is likey to be small, as discussed in the introduction to this paper.}
Using a 580 ks exposure, an all-Crab polarisation fraction of (28$\pm$6)\% was measured using the {\it INTEGRAL} SPI instrument for a polarisation angle of (117$\pm$9)$^\circ$ (130-440~keV)~\citep[][]{Chauvin2013}. 
The {\it INTEGRAL} IBIS instrument was used by~\citet{Forot2008} to determine an all-Crab polarisation fraction of (47$^{+19}_{-13})$\% for a polarisation angle of (100$\pm$11)$^\circ$ (200-800~keV). 
The precision of our data and that from {\it INTEGRAL} does not currently permit a meaningful analysis of the energy dependence of the polarisation fraction and angle. 

The maximum allowed polarisation fraction for synchrotron emission in a uniform magnetic field is $\sim$75\%~\citep[][]{Lyutikov2003}, assuming a relevant parent electron spectral index of $\sim$3~\citep[][]{Kuiper2001}.
Our measurement is $>$5 standard deviations from this maximum, with the central value suggesting an ordered environment at the emission site.
The measured polarisation angle is consistent within 1.6 standard deviations from the {\color{black} inferred} direction of the pulsar spin axis reported by~\citet{Ng2004} in the optical regime. 
This alignment suggests that polarised hard X-ray emission originates close to the pulsar. Possible emission sites have been identified through 
a detailed study at optical wavelengths by~\citet{Moran2013} which shows that the polarisation of wisps, the inner knot and torus are well aligned to the pulsar spin axis 
and have $\sim$50\%, polarisation fraction. 

A second high latitude long duration balloon flight will be conducted in July 2016. Based on experience from the 
2013 flight, we will improve the polarimeter design with particular focus on improving the $M_{100}$ and background reduction
methods~\citep[][]{Chauvin2015b}. This will allow the contribution from the nebula to be isolated through pulse phase selections.

%%%%%%%%%%%%%%%%%%%%%%%%%%%%%%%%%%%%%%%%%%%%%%%%%%
\section*{Acknowledgements}

The {\it PoGOLite} Collaboration acknowledges funding received from The Swedish National Space Board, The Knut and Alice Wallenberg Foundation, The Swedish Research Council, The G\"{o}ran Gustafsson Foundation, SLAC/KIPAC (Stanford University), JAXA, and
%Hiroshima University
JSPS KAKENHI (grant numbers 16340055, 18340052, 14079102, 14GS0211, 24103002, 18740154, 22540245, 14079207, 21740186, 23740193, 25302003).
The SSC Esrange Space Centre is thanked for their support during the {\it PoGOLite Pathfinder} flight campaign. 
All past members of the {\it PoGOLite} Collaboration not listed as authors on this paper are thanked for their important contributions to the development of the project. 

%%%%%%%%%%%%%%%%%%%%%%%%%%%%%%%%%%%%%%%%%%%%%%%%%%



%\bibliographystyle{mnras}

\begin{thebibliography}{99}

%\bibitem{Fermi} A. A.  Abdo, et al., ApJ 708 (2010) 1254.
\bibitem[\protect\citeauthoryear{Abdo et al.}{2010}]{Abdo2010}
Abdo A. ~A., et al., 2010, ApJ, 708, 1254.

%\bibitem{SPIall} M. Chauvin, et al., ApJ 769:137 (2013) 1.
\bibitem[\protect\citeauthoryear{Chauvin et al.}{2013}]{Chauvin2013}
Chauvin M., et al., 2013, ApJ, 769:137, 1.

%\bibitem{payload} M.~Chauvin, et~al.,
%\newblock {Submitted to Experimental Astronomy (2015)}.
\bibitem[\protect\citeauthoryear{Chauvin et al.}{2015a}]{Chauvin2015a}
Chauvin M., et al., 2015a, Experimental Astronomy (accepted for publication). arXiv:1508.03345.

\bibitem[\protect\citeauthoryear{Chauvin et al.}{2015b}]{Chauvin2015b}
Chauvin M, et al., 2015b, Astroparticle Physics (in preparation).

%\bibitem{calibration} M.~Chauvin, et~al.,
%\newblock {Astroparticle Physics (2015), accepted for publication. arXiv:1505.08093}.
\bibitem[\protect\citeauthoryear{Chauvin et al.}{2016}]{Chauvin2016}
Chauvin M., et al., 2016, Astroparticle Physics, 72, 1. 
%doi:10.1016/j.astropartphys.2015.05.003.

%outer gap
\bibitem[\protect\citeauthoryear{Cheng}{2000}]{Cheng2000} 
Cheng K.S., et al., ApJ, 537, 964.

% polar cap
\bibitem[\protect\citeauthoryear{Daugherty et al.}{1996}]{Daugherty1996}
Daugherty J.K., Harding, A.K.,1996, ApJ, 458, 278.

%\bibitem{SPIoff} A. J. Dean, et al., Science 321 (2008) 1183.
\bibitem[\protect\citeauthoryear{Dean et al.}{2008}]{Dean2008}
Dean A.~J., et al., 2008, Science, 321, 1183.

% Caustic
\bibitem[\protect\citeauthoryear{Dyks et al.}{2004}]{Dyks2004}
Dyks J., et al., 2004, ApJ, 606 1125. 

%\bibitem{IBIS} M. Forot, et al., ApJ 688 (2008) L29.
\bibitem[\protect\citeauthoryear{Forot et al.}{2008}]{Forot2008}
Forot M., et al., 2008, ApJ, 688, L29.

% polar cap
%\bibitem[\protect\citeauthoryear{Daugherty et al.}{1982}]{Daugherty1982}
%Daugherty J.K., Harding, A.K.,1982, ApJ, 252, 337.
% polar cap
\bibitem[\protect\citeauthoryear{Harding et al.}{1998}]{Harding1998}
Harding A.K., Muslimov G.A.,1998, ApJ, 508, 328. 

%General Crab
\bibitem[\protect\citeauthoryear{Hester}{2008}]{Hester2008}
Hester J. J., 2008, ARA\&A, 46, 127.

%\bibitem{Kislat201545} F.~Kislat, et~al.,
%\newblock {Astroparticle Physics 68 (2015) 45}.
\bibitem[\protect\citeauthoryear{Kislat et al.}{2015}]{Kislat2015}
Kislat F., et al., 2015, Astroparticle Physics, 68, 45.

\bibitem[\protect\citeauthoryear{Kokubun et al.}{2007}]{Kokubun2007}
Kokubun M., et al., 2007, Publications of the Astronomical Society of Japan, 59, S53.

\bibitem[\protect\citeauthoryear{Krawczynski et al.}{2011}]{Krawczynski2011}
Krawczynski H., et al., 2011, Astroparticle Physics, 34, 550.

%\bibitem{SwiftBAT2013} H.~A. Krimm, et~al.,
%\newblock {ApJ Supp 209 (2013) 14}.
\bibitem[\protect\citeauthoryear{Krimm et al.}{2013}]{Krimm2013}
Krimm H.~A., et al., 2013, ApJ Suppl, 209, 14.

%\bibitem{spectral} L.~Kuiper, et al., A\&A 378 (2001) 918.
\bibitem[\protect\citeauthoryear{Kuiper et al.}{2001}]{Kuiper2001}
Kuiper L., et al., 2001, A\&A, 378, 918.

\bibitem[\protect\citeauthoryear{Lei et al.}{1997}]{Lei1997}
Lei F., et al., 1997, Space Sci. Rev., 82, 309.

\bibitem[\protect\citeauthoryear{Lyutikov et al.}{2003}]{Lyutikov2003}
Lyutikov M., et al., 2003, ApJ, 597, 998.

%\bibitem{Maier2014} D.~Maier, et~al.,
%\newblock {Astronomical Society of the Pacific 126 (2014) 459}.
\bibitem[\protect\citeauthoryear{Maier et al.}{2014}]{Maier2014}
Maier D., et al., 2014, Astronomical Society of the Pacific, 126, 459.

%\bibitem{methodology} V.~Mikhalev, et~al.,
%\newblock {In preperation}.
\bibitem[\protect\citeauthoryear{Mikhalev et al.}{2015}]{Mikhalev2015}
Mikhalev V., et al., 2015, A\&A (in preparation).

\bibitem[\protect\citeauthoryear{Moran et al.}{2013}]{Moran2013}
Moran P., et al., 2013, arXiv:1302.3622v1.

\bibitem[\protect\citeauthoryear{Novick et al.}{1972}]{Novick1972}
Novick R., et al., 1972, ApJ, 174, L1.

%\bibitem{Ng} C. -Y. Ng and R. W. Romani, ApJ 601 (2004) 479.
\bibitem[\protect\citeauthoryear{Ng et al.}{2004}]{Ng2004}
Ng C.~-Y., Romani R.~W., 2004, ApJ, 601, 479.

% outer gap
%\bibitem[\protect\citeauthoryear{Cheng et al.}{1986}]{Cheng1986a}
%Cheng K.S., Ho C., Ruderman M.A.,1986, ApJ, 300, 500. 
%outer gap
%\bibitem[\protect\citeauthoryear{Cheng et al.}{1986}]{Cheng1986b}
%Cheng K.S., Ho C., Ruderman M.A,1986, ApJ, 300, 522.
%outer gap

\bibitem[\protect\citeauthoryear{Quinn}{2012}]{Quinn2012}
Quinn J.L., 2012, A\&A, 538, A65.

\bibitem[\protect\citeauthoryear{Romani}{1996}]{Romani1996} 
Romani R.W.,1996, ApJ, 470, 469. 

\bibitem[\protect\citeauthoryear{Satori et al.}{1967}]{Satori1967}
Satori L., Morrison P., 1967, ApJ, 150 385. 

%Optical
\bibitem[\protect\citeauthoryear{S\l{}owikowska et al.}{2009}]{Slowikowska2009}
S\l{}owikowska A., et al., 2009, MNRAS, 397, 103.

%XIPE
\bibitem[\protect\citeauthoryear{Soffitta et al.}{2013}]{Soffitta2013}
Soffitta P., et al., Experimental Astronomy, 2013, 36 (3), 523.

%Curvature
\bibitem[\protect\citeauthoryear{Sturrock et al.}{1975}]{Sturrock1975}
Sturrock P.A., et al., 1975, ApJ,196, 73.

% HXD
\bibitem[\protect\citeauthoryear{Takahashi et al.}{2007}]{Takahashi2007}
Takahashi T., et al., 2007, Publications of the Astronomical Society of Japan, 59, S35.

% Crab spectrum
%\bibitem[\protect\citeauthoryear{Toor et al.}{1974}]{Toor1974}
%Toor A., et al., 1974, Astron. J., 995.

%\bibitem{weisskopf1976} M. C. Weisskopf, et al., ApJ 208 (1976) 125. 
\bibitem[\protect\citeauthoryear{Weisskopf et al.}{1976}]{Weisskopf1976}
Weisskopf M.~C., et al., 1976, ApJ, 208, 125. 

\bibitem[\protect\citeauthoryear{Weisskopf et al.}{2008}]{Weisskopf2008}
Weisskopf M.~C., et al., 2008, Proceedings of SPIE, 7022, 70111I. 

%\bibitem{Weisskopf2010SPIE} M.~C. Weisskopf, et~al.,
%\newblock {Proceedings of SPIE 7732 (2010) 77320E}
\bibitem[\protect\citeauthoryear{Weisskopf et al.}{2010}]{Weisskopf2010}
Weisskopf M.~C., et al., 2010, Proceedings of SPIE, 7732, 77320E.

\end{thebibliography}

% FORMAT:


%%%%%%%%%%%%%%%%%%%%%%%%%%%%%%%%%%%%%%%%%%%%%%%%%%

%%%%%%%%%%%%%%%%% APPENDICES %%%%%%%%%%%%%%%%%%%%%

%\appendix


%%%%%%%%%%%%%%%%%%%%%%%%%%%%%%%%%%%%%%%%%%%%%%%%%%

% Don't change these lines
\bsp	% typesetting comment
\label{lastpage}
\end{document}